\documentclass[rmp,twocolumn,floatfix]{revtex4}
\usepackage{graphicx,amsmath,amssymb,url,txfonts}

\begin{document}

\title{Multiscaling in the \textit{YX} model of networks}

\author{Petter Holme}
\author{Zhi-Xi Wu}
\author{Petter Minnhagen}

\affiliation{Department of Physics, Ume{\aa} University, 901~87 Ume{\aa}, Sweden}

\begin{abstract}
We investigate a Hamiltonian model of networks. The model is a mirror formulation of the \textit{XY} model (hence the name) --- instead letting the \textit{XY} spins vary, keeping the coupling topology static, we keep the spins conserved and sample different underlying networks. Our numerical simulations show complex scaling behaviors, but no finite-temperature critical behavior. The ground state and low-order excitations for sparse, finite graphs is a fragmented set of isolated network clusters. Configurations of higher energy are typically more connected. The connected networks of lowest energy are stretched out giving the network large average distances. For the finite sizes we investigate there are three regions --- a low-energy regime of fragmented networks, and intermediate regime of stretched-out networks, and a high-energy regime of compact, disordered topologies. Scaling up the system size, the borders between these regimes approach zero temperature algebraically, but different network structural quantities approach their $T=0$-values with different exponents.
\end{abstract}

\maketitle

\section{Introduction}

The \textit{XY} model is one of the classical and most versatile spin models of statistical mechanics. It has been used in many physical contexts, from superconductors~\cite{suzuki:rev} to lattice gauge theory of quantum chromodynamics~\cite{meyer:qcd}, and further on applied to emergent phenomena in systems like bird flocks~\cite{xybird} and parallel discrete-time simulations~\cite{korn:sup}. In general, the \textit{XY} model describes a system of $N$ pairwise interacting units. Let the interaction be represented by a graph $G=(V,E)$ where $V$ is the set of units, or vertices, and $E$ is the set of \textit{edges} (pairs of coupled units) Each unit $i\in V$ is characterized by a real number $\theta_i$ called the \textit{spin} of $i$. The probability of a certain combination $G$ and $\{\theta_i\}_{i\in V}$ is proportional to $\exp\, (-H/T)$ where the parameter $T$ is called temperature (parameterizing the disorder of the system), and $H$ is the Hamiltonian
\begin{equation}\label{eq:hamil}
H=-\sum_{(i,j)\in E} \cos\,(\theta_i-\theta_j) 
\end{equation}
--- a function that in physical systems represents the energy of a configuration. (Note that, in our formalism, temperature and energy are dimensionless.) From the symmetry of the cosine function in Eq.~(\ref{eq:hamil}), we see that the values of $\theta_i$ only matters modulo $2\pi$. For this reason is $\theta_i$ commonly restricted to the interval $(0,2\pi]$.

Traditionally, studies of the \textit{XY} model take $G$ as a lattice graph --- a graph that can be drawn as a lattice (a discrete subgroup spanning the vector space $\mathbb{R}^d$, where $d$ is the dimension, a natural number) --- and considered fixed, while $\{\theta_i\}_{i\in V}$ is the object of study. (The most important result is perhaps that for a two-dimensional lattice structure the \textit{XY} model undergoes a peculiar phase transition, the Kosterlitz--Thouless transition~\cite{kt:trans} between a disordered phase and a phase with algebraic spin correlations.) Sometimes random graph ensembles are used~\cite{our:swxy,katya:swxy,xy:sf}, sometimes regular topologies that are not mathematical lattices~\cite{seungi:xy}, but usually $G$ is static (the only exception we are aware of is Ref.~\cite{xyflu}, where the \textit{XY} model is simulated on a graph being rewired without any bias).  In this paper we turn the situation around --- we fix $\{\theta_i\}_{i\in V}$ and let $G$ be only restricted by $N$, the number of edges $M$ and that the graph is simple (i.e., that there are no multiple edges or self-edges). The spins are, like common practice when initializing the \textit{XY} model for Monte Carlo simulations, drawn with uniform probability. Our model is a conceptual mirror image of the \textit{XY} model --- hence we call it the \textit{YX} model.

The \textit{YX} model defines an ensemble of graphs and is thus more related to the recent works on models of complex networks~\cite{ba:rev,mejn:rev,doromen:book}. Authors have studied general classes of graphs sampled with a Boltzmann-like probability --- so called exponential random graphs~\cite{erg:rev}, Markov graphs~\cite{strauss_frank}, p-star models~\cite{pstar1} or statistical-mechanics models~\cite{park:sm} depending on the background of the author. These models are used for sampling networks with some prescribed structures; they are thus not microscopic, or mechanistic, models and usually not analyzed in terms of emergent properties in the $N\rightarrow\infty$ (``thermodynamic'') limit. The \textit{YX} model is also related to hidden variable models~\cite{cald:go,motter:sn,kahng:qc} where some variable (in this case $\theta_i$) assigned to the vertices is affecting their position in the network.

The main reason of investigating the \textit{YX} model is somewhat academic --- what happens when we update the connections, rather than the spins, in a simple statistical-physics model?  A token that the \textit{YX} model has a rich and interesting behavior is that the ground state is frustrated (i.e., the terms, $-\cos\,(\theta_i-\theta_j)$ in the sum of Eq.~(\ref{eq:hamil}) is not minimal ($-1$) for all pairs $(i,j)\in E$. If we make a similar modified Ising, model or $q$-state Potts model, if the number of edges is small enough (smaller than the number of edges in a graph where all nodes of the same spin have an edge between them), the system is not frustrated. Such models will most likely have a second-order phase transition between segregated and mixed states and might be interesting in their own rights. In the \textit{YX} model it is hard to guess the outcome, even the ground state, \textit{a priori}. 

In the rest of the paper we will present the simulation scheme in more detail and analyze the size scaling of network structural quantities.

\section{Simulations}

We simulate the \textit{YX} model using Metropolis Monte Carlo sampling~\cite{mejn:statmech}. For one update \textit{step}, we choose three distinct vertices $i$, $j$ and $k$ such that $(i,j)\in E$ but $(i,k)\notin E$ at random. Let $E'$ be $E$ with $(i,j)$ replaced by $(i,k)$, then we accept the change (replacing $E$ by $E'$) if $H(E)\geq H(E')$; or, if $H(E)< H(E')$, with a probability proportional to
\begin{equation}\label{eq:metro}
\exp\,\left (\frac{H(E)-H(E')}{T}\right) .
\end{equation}
Let, furthermore, $N$ update steps comprise one \textit{sweep}.

The configuration space of the \textit{YX} model is, as will be discussed later, probably not very rugged. Still, to be on the safe side, we use the Exchange Monte Carlo method~\cite{xmc} capable of, even for glassy systems with many local energy minima, sampling the configuration space evenly in a limited time. In Exchange Monte Carlo an even number of systems are simulated in parallel for a sequence of temperatures. After some interval, with a probability dependent on the current configuration two systems at adjacent temperatures are exchanged, so that one system moves up in temperature, the other one down. In our simulations we test for an exchange every $10^5$'th sweep. We measure network quantities with the same frequency as the tests for exchanges. Before we start measuring, we run $10^7$ sweeps for the system to reach equilibrium (which is roughly $10$ times longer than it takes for all quantities for all sizes and temperatures to converge). We use $100$ measurements to calculate intermediate averages. This procedure is then repeated for $100$ random initial conditions and the averages and standard errors of the intermediate averages are the values we present below.

Unless otherwise stated we will use $M=2N$. We chose an exponential set of $20$ temperatures per system size selected in a preliminary study to capture the most interesting region.

\section{Numerical results}

\begin{figure}
\includegraphics[width=\linewidth]{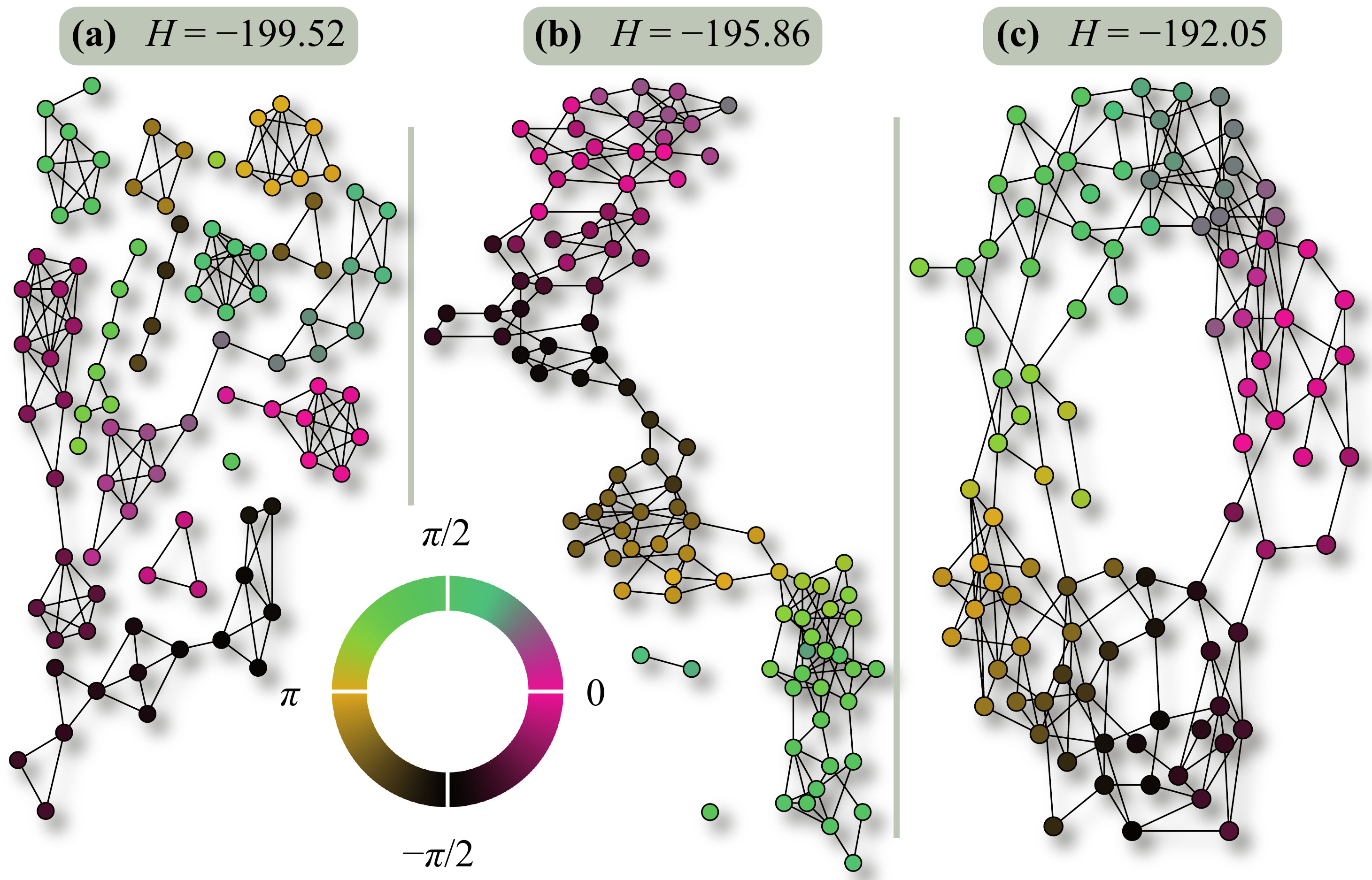}
  \caption{Three low-energy configurations of the \textit{YX} model with $100$ vertices. }
  \label{fig:conf}
\end{figure}

\subsection{Low-energy configurations}

What is the ground state configuration of the \textit{YX} model? First consider a simpler example, --- suppose $N$ is a multiple of three, $N=M$  and the angles $\{\theta_i\}_{i\in V}$ are evenly spread out over the circle, i.e.\ $\theta_i=2\pi i / N$. The minimal distance between any pair of $\theta_i$ is the distance $2\pi / N$ between two adjacent vertices along the circle. There are exact $M$ such pairs, so the ground state is $E=\{(1,2), (2,3), \cdots , (N-1,N), (N,1)\}$. Let $f=M+H$ be the \textit{frustration} --- how far from the lowest energy of a model where $\{\theta_i\}_{i\in V}$ is unconstrained, the configuration is. For the state of circularly coupled edges, we have (by Taylor expansion, $\cos \phi \approx 1 - \phi^2/2$ for small $\phi$) $f=M - M\cos(2\pi/N)\approx  2M\pi^2/N^2=2\pi^2/N$ for large $N$. Another low-energy configuration would be to couple nearby vertices into $N/3$ triangles --- $E=\{(1,2), (2,3), (3, 1), \cdots , (N-2,N-1), (N-1,N), (N,N-2)\}$. The frustration in this case is twice as large as the circular configuration. If we change the system above, so $\theta_1=2\pi  / N+\delta_\theta$ (for some small perturbation $0<\delta_\theta<\pi/N$) and the rest is the same, then the frustration increase with $\delta_\theta^2$ for the circular configuration, while it decreases by $6\pi^2\delta_\theta/N-\delta_\theta^2$ for the configuration of isolated triangles. This example can be fairly straightforwardly generalized to higher edge densities. It suggests that fragmented configurations benefit from an irregular distribution of angles, while circular distributions should give the lowest energies at evenly distributed angles. Since we sample $\{\theta_i\}_{i\in V}$ by uniform randomness, in the $N\rightarrow\infty$ limit the ground state should be a circulant (a graph where vertices are connected to their nearest neighbors on a circle). The fragmented states can utilize the gaps in the distribution of drawn $\theta_i$-values by omitting edges across such gaps. For finite sizes, it could happen that the ground state is fragmented rather than a circulant. Indeed, this is what we see in our simulations.

In Fig.~\ref{fig:conf} we show three low-energy configurations for a small system size. The lowest energy state, with $f=0.48$, is fragmented; the other two, with $f=4.14$ and $7.95$ are more elongated, closer to circulants. As we will see, for finite sizes, fragmented states like Fig.~\ref{fig:conf}(a) have the lowest energies while the circulants have larger entropies making configurations like Figs.~\ref{fig:conf}(b) and (c)  dominant at higher temperatures. In the large temperature limit, the networks are Erd\H{o}s--R\'{e}nyi random graphs~\cite{ba:rev,mejn:rev,doromen:book}.

\begin{figure*}
\includegraphics[width=0.7\linewidth]{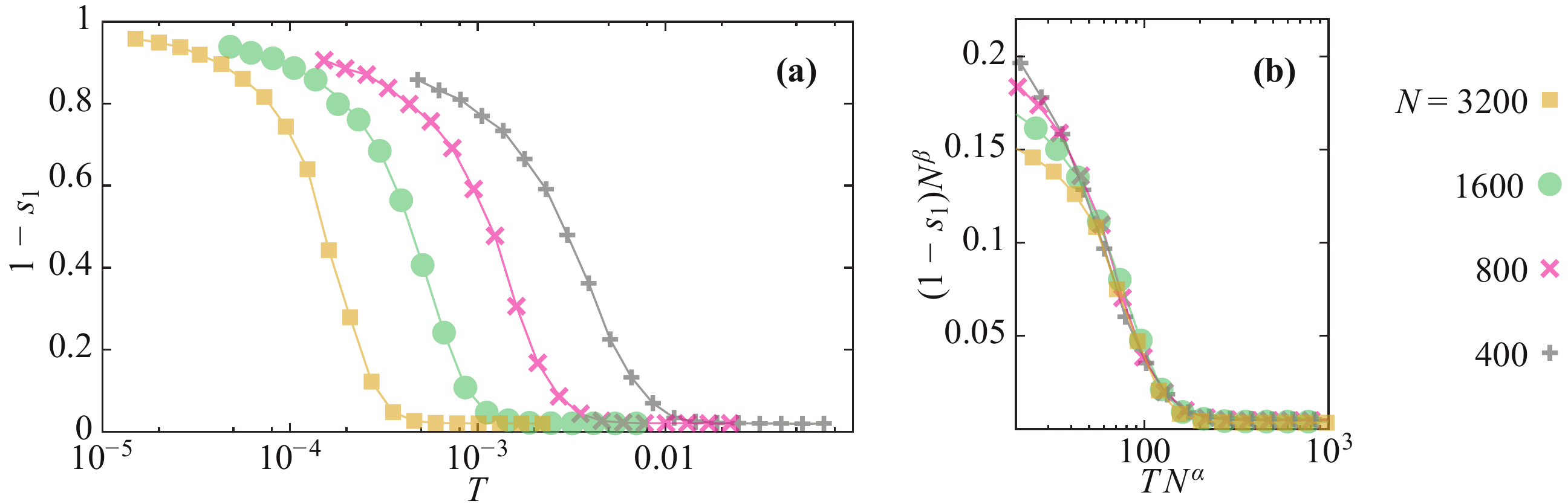}
  \caption{ The fraction of vertices not part of the largest connected cluster. In (a) we show the raw data as a function of temperature for four different system sizes. In (b) we determine the scaling exponents for low temperatures to $\alpha=1.61\pm 0.05$ and $\beta=0.22\pm 0.03$.  Standard errors are smaller than the symbol size and omitted for clarity. Lines are guides for the eyes. }
  \label{fig:s1}
\end{figure*}

\begin{figure}
\includegraphics[width=0.92\linewidth]{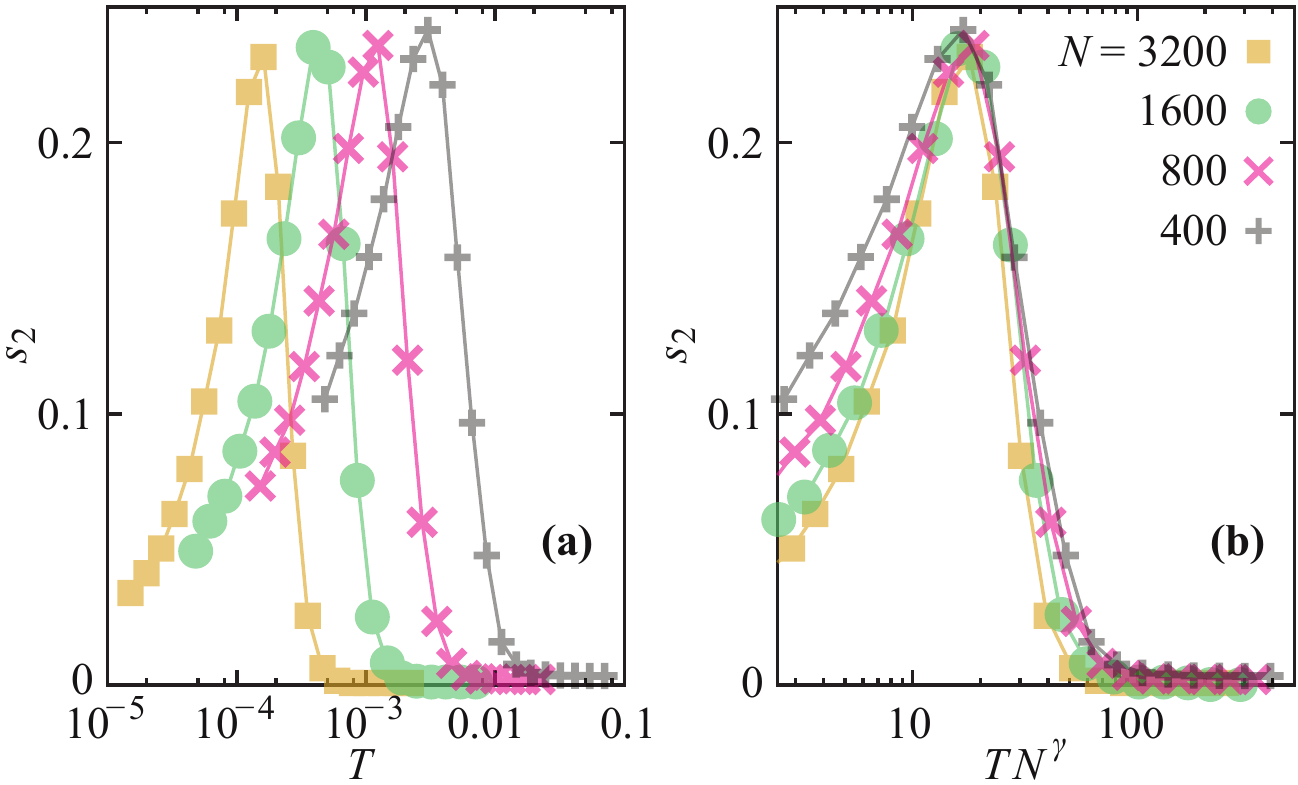}
  \caption{ The fraction of the second largest cluster. (a) displays the raw data of $s_2$ as a function of temperature. In (b) we determine the scaling factor $\gamma=1.44\pm 0.02$. Standard errors are smaller than the symbol size and omitted for clarity. Lines are guides for the eyes.}
  \label{fig:s2}
\end{figure}

\subsection{Cluster sizes}

Now we turn to the quantitative results. In Fig.~\ref{fig:s1} we plot the relative fraction of the vertices not a part in the largest cluster, $1-s_1$. Compared with spin models like the \textit{XY} model, the temperatures where the interesting behavior occurs is much lower. The interesting behavior --- the fragmentation of the network --- is best monitored in a logarithmic temperature scale. The onset of fragmentation approaches (as expected from the discussion above), zero as the system size increases.  In Fig.~\ref{fig:s1}(b) we show that for $\alpha=1.61\pm 0.05$ and $\beta=0.22\pm 0.03$ we have
\begin{equation}
1-s_1\approx N^{-\beta}F_+(TN^{\alpha}) ,
\end{equation}
for a smooth function $F$, and $TN^{1.61}\gtrsim 80$. The onset of fragmentation thus moves to zero like $N^{-\alpha}$. Note that this scaling relation is itself not indicative of critical behavior in the rescaled temperature $TN^{\alpha}$. The observation that the scaling region is larger for the smaller than the larger system sizes, which is also not supporting an emergent critical behavior in $s_1$.

An unexpected phenomenon in the \textit{XY} model is that there seems to be no universality in this size scaling of the network-structural quantities. An example of this is seen in Fig.~\ref{fig:s2} where we plot the temperature dependence of the relative size of the second largest component $s_2$. In many systems with a fragmentation phase transition, like percolation or network models of segregation~\cite{our:coevo,zanet:coevo}, $s_2$ or $s_2/s_1$ can be used to characterize the critical behavior. Also in our case, $s_2$ gives a strong signal of the low-temperature fragmentation. The heights of the peaks are quite independent of system size. The location of the peak scales to zero like $T\sim N^{-\gamma}$, $\gamma=1.44\pm 0.02$. This means that the peak of $s_2$ goes to zero slower than the onset of fragmentation as seen in $s_1$.

\begin{figure}
\includegraphics[width=\linewidth]{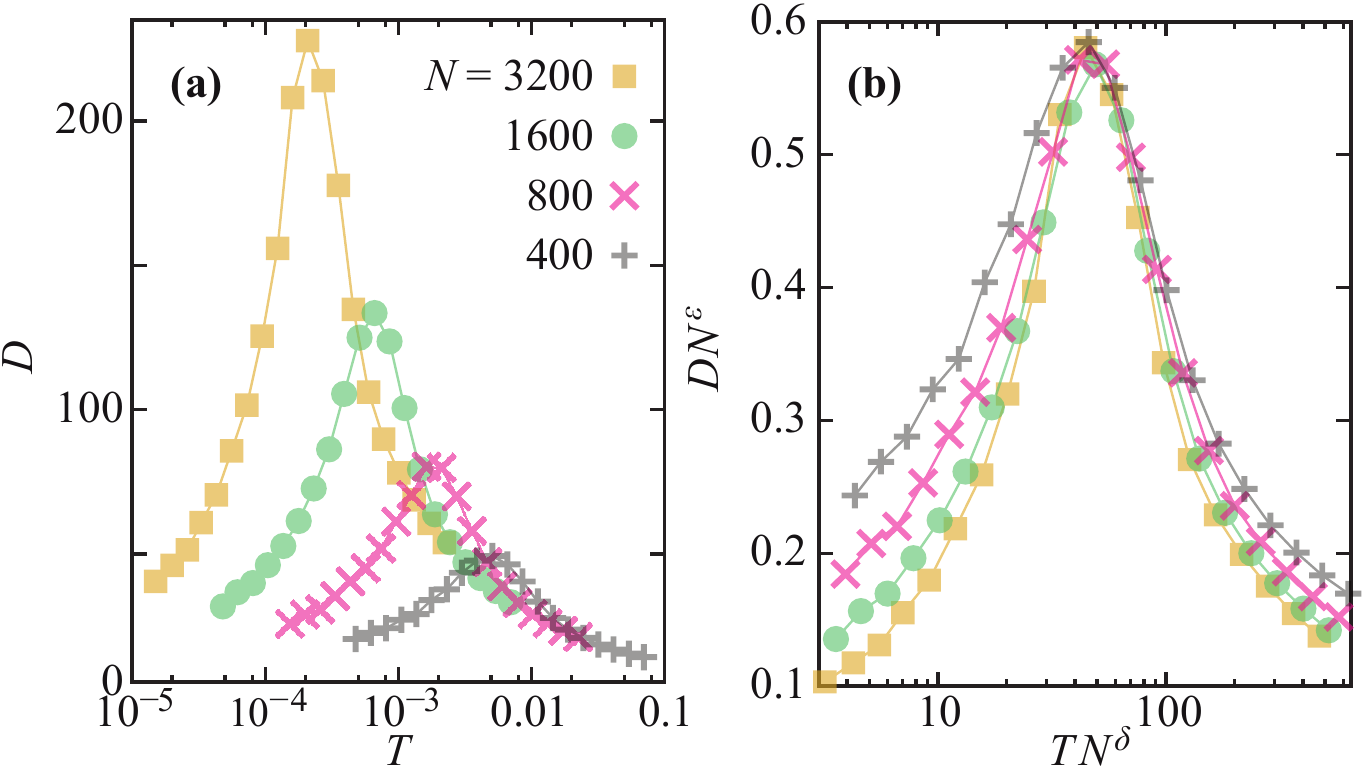}
  \caption{ Scaling of the diameter of the largest connected cluster $D$ as a function of temperature (a), and the determination of $D$'s scaling parameters $\delta$ and $\epsilon$ (b). We find $\delta = 1.52\pm 0.02Ñ$ and $\epsilon = -0.74\pm 0.02$. Standard errors are smaller than the symbol size and omitted for clarity. Lines are guides for the eyes.}
  \label{fig:dia}
\end{figure}

\subsection{Diameter}

The \textit{distance} between $i$ and $j$ is the number of edges in the shortest path between the two nodes. The largest distance in a connected subgraph is the \textit{diameter} $D$ of that subgraph. If the picture from Fig.~\ref{fig:conf} holds --- that at intermediate temperatures are dominated by stretched out configurations, and the networks, at lower temperatures, are fragmented --- then the diameter of the largest connected component should have a peak at intermediate temperatures. In Fig.~\ref{fig:dia}(a) we plot the diameter as a function of temperature. Indeed there is a peak moving to zero like $T\sim N^{-\delta}$, $\delta = 1.52\pm 0.03$, and increasing in size as $N^{-\epsilon}$ with $\epsilon = -0.74\pm 0.02$. For the sizes we test, the peak in $D$ occurs at slightly higher temperatures than the peak in $s_2$. It also scales slightly faster than the $s_2$ peak towards zero. The fact that the peak of $D$ scales sublinearly reflects that the intermediate region is a mix of configurations, not all stretched out like Fig.~\ref{fig:conf}(b) and (c). This is of course a fundamental aspect of Hamiltonian models --- all configurations have a finite chance of appearing at all temperatures, but their frequencies vary with the temperature. The fact that the peak value of $D$ diverges with $N$ supports the picture of circulant ground states (in the $N\rightarrow\infty$). The peak in Fig.~\ref{fig:dia}(b) becomes sharper with larger system sizes, meaning that the increase as $TN^\delta$ is lowered gets more dramatic. This observation is in concordance with a $T=0$ phase transition. Our data does, however, not give a very strong support for a critical scaling of $D$ at the sizes we investigate.

\begin{figure}
\includegraphics[width=0.95\linewidth]{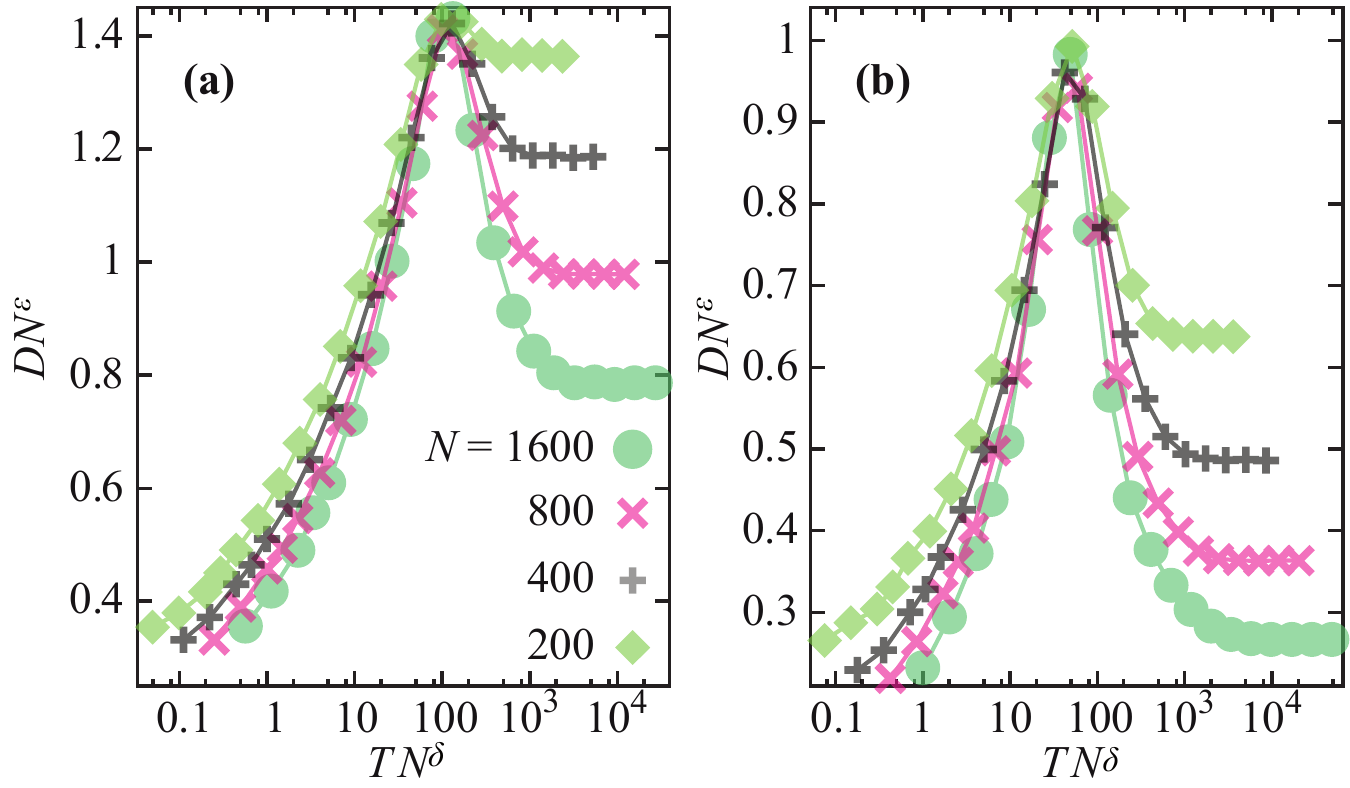}
  \caption{ Figures corresponding to Fig.~\ref{fig:dia}(b) for $3N=2M$ (a) and $N=M$ (b). The scaling exponents for (a) are determined to $\delta = 1.17\pm 0.10$ and $\epsilon = -0.52\pm 0.02$; for (b) they are $\delta = 1.25\pm 0.10$ and $\epsilon = -0.61\pm 0.02$. Standard errors are smaller than the symbol size and omitted for clarity. Lines are guides for the eyes.}
  \label{fig:dia_other_dens}
\end{figure}

\subsection{Density dependence}

We perform most of our analysis for $M/N=2$ but will briefly touch on how the scaling depends on the density of edges. As our example we choose the scaling analysis of the diameter seen in Fig.~\ref{fig:dia}(b), but all the other scaling plots give the same conclusion, see  Fig.~\ref{fig:dia_other_dens}. Both the networks with $M/N=2/3$ (a) and $M/N=1$ (b) have peaks of the diameter scaling to zero with increasing sizes. The peak is less marked in the sparser graphs in (a) which is natural since they are closer to the fragmentation threshold at high temperatures ($M/N=1/2$). The exponents are also $N$-dependent, with $\delta$ decreasing with decreasing edge density, and $\epsilon$ increasing (from more to less negative values) as the density becomes larger. If the density is larger,  e.g.\ $M/N=4$, the scaling we plot, at least for the sizes we measure break down (so that there is no $\delta$ such that $D$ peaks at the same $TN^\delta$ and there is no $\epsilon$  such that the maximal $TN^\epsilon$ is the same for all $N$).

\begin{figure}
\includegraphics[width=0.97\linewidth]{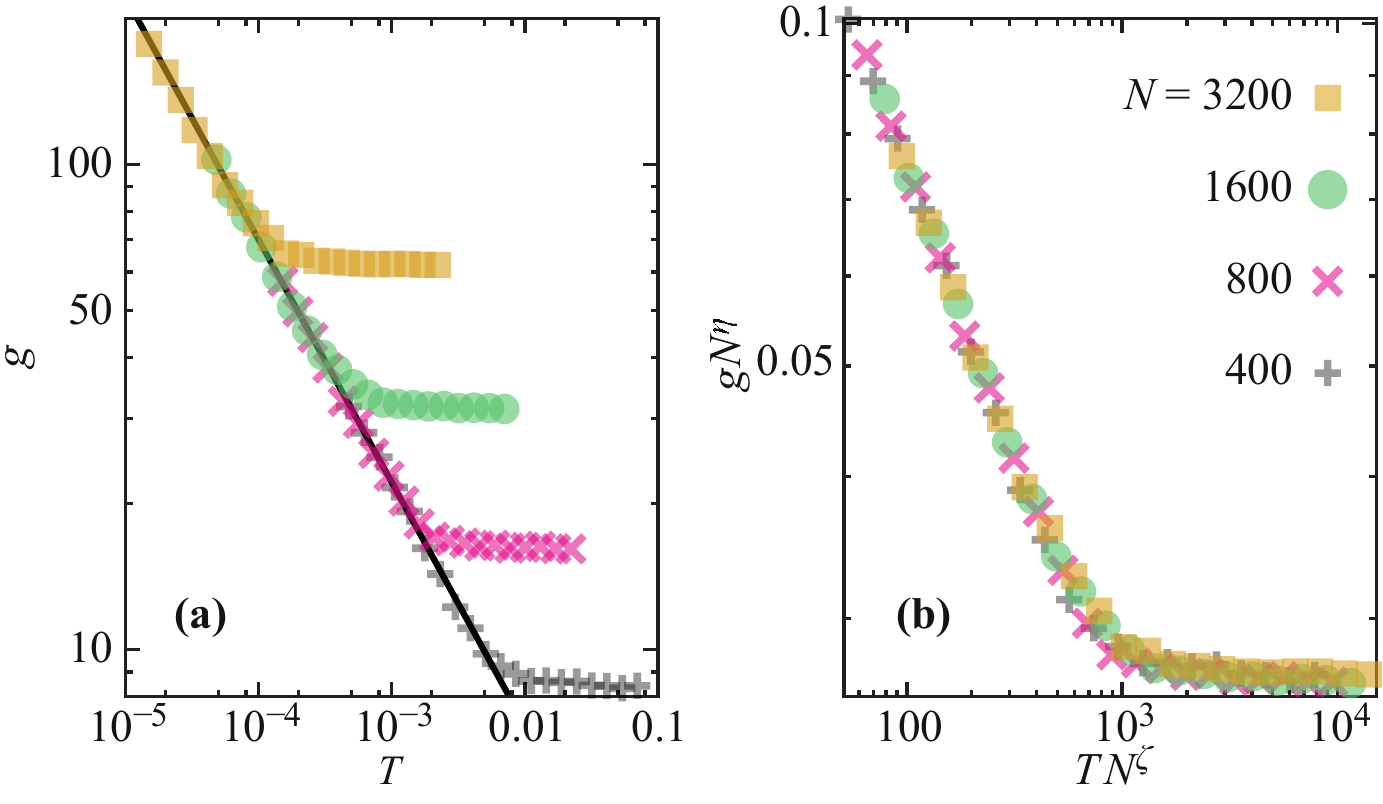}
  \caption{ The number of fragments as a function of temperature (a). The solid line is proportional to $T^{-1/2}$. (b) shows the determination of the scaling exponents $\zeta=1.94\pm 0.07$ and $\eta=0.96\pm 0.03$. Standard errors are smaller than the symbol size and omitted for clarity. }
  \label{fig:nfrag}
\end{figure}

\subsection{Number of isolated subgraphs}

Another feature characterizing the fragmentation is the number of connected subgraphs $g$. In Fig.~\ref{fig:nfrag}(a) we plot this quantity as a function of temperature. The fragmented and unfragmented states are rather conspicuously separated and in the low-temperature region $g=\Gamma T^{-1/2}$ with the same constant $\Gamma$ for all system sizes. From Fig.~\ref{fig:nfrag}(b) we see that
\begin{equation}\label{eq:nfrag_scale}
g=N^{-\eta}G(TN^\zeta) ,
\end{equation}
where $\zeta=1.94\pm 0.07$, $\eta=0.96\pm 0.03$, and $G$ is the function giving the shape seen in Fig.~\ref{fig:nfrag}(b). In traditional statistical physics, for exactly solvable models, the scaling exponents often have rational values. The $g$-scaling is consistent with a simple picture: if $TN^2\lesssim 800$ we have $g \sim T^{-1/2}$; if $TN^2\gtrsim 1600$ we have $g \sim N$.  To explain the value of $\zeta$, we note that at the energy close to the ground state is governed by angular differences scaling like $1/N$. Hence $N^{-2}$ (by the leading, square term of the frustration) gives a local energy scale, which in analogy to van der Waals' law of corresponding states implies that $T/N^{-2}$ is the fundamental quantity for the fragmentation, i.e. that $\zeta=2$. The large $TN^\zeta$ scaling, $g\sim N$, can be understood from the Erd\H{o}s--R\'{e}nyi model that is the high temperature limit of the \textit{YX} model and has the same scaling behavior. The low $TN^\zeta$ scaling, $g\sim T^{-1/2}$, is related to the state with fragmented dense clusters. Assume that all clusters are of similar sizes, then there are $N/g$ vertices in each cluster. The number of edges in a dense cluster scales like $(N/g)^2$, and thus the total number of edges like $M=g(N/g)^2=N^2/g$. Since we assume sparse graphs, $M\sim N$, the number of clusters in such a graph scales like $g\sim N$. By the same argument as above, that $TN^2$ is a fundamental quantity, we get $g \sim T^{-1/2}$.

\begin{figure*}
\includegraphics[width=\linewidth]{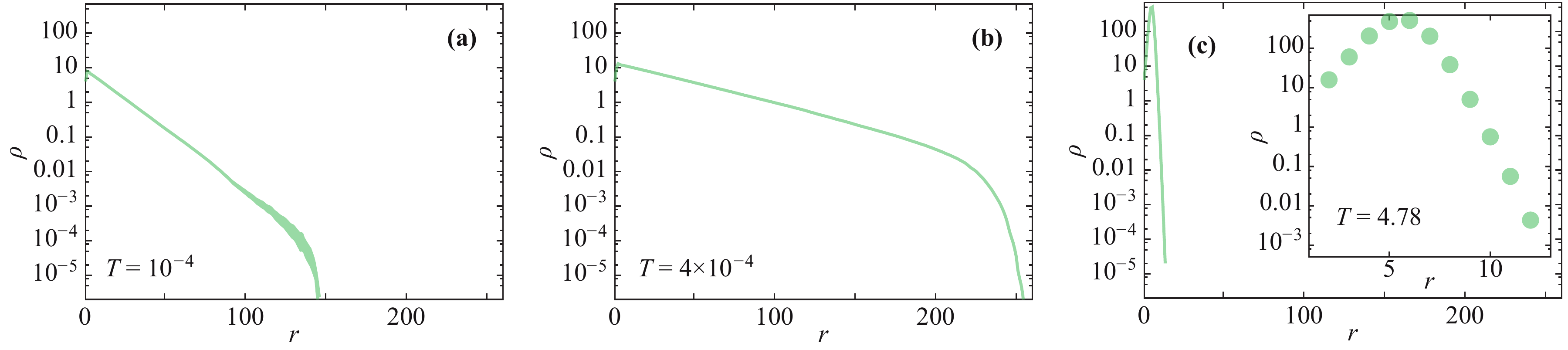}
  \caption{ The average density of vertices at a distance $r$ from a vertex for three different temperatures (indicated in the panels) and $N=1600$. To facilitate comparison, we let the panels cover the same ranges. The widths of the lines are at least the standard error. The inset in panel (c) is a blow-up.}
  \label{fig:dens}
\end{figure*}

\begin{figure*}
\includegraphics[width=\linewidth]{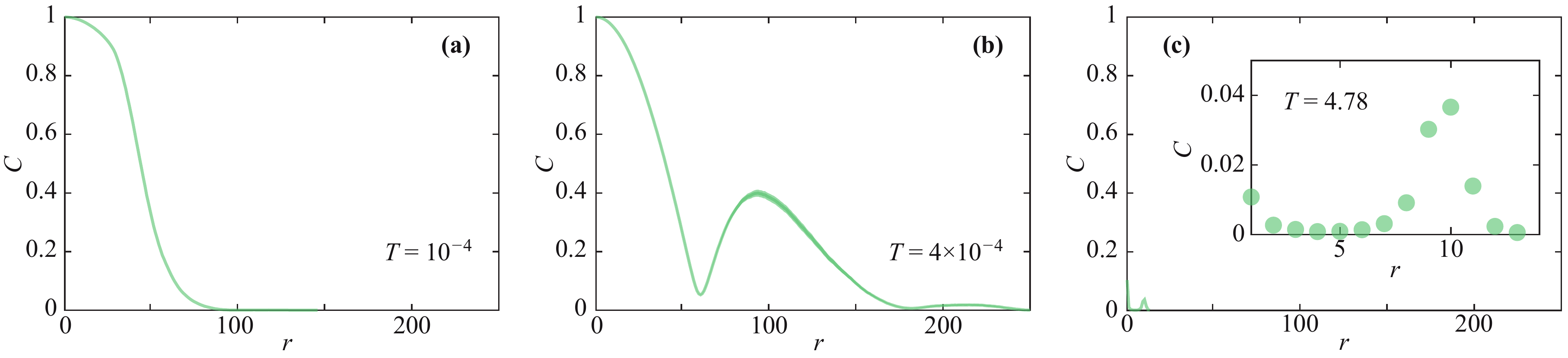}
  \caption{ The correlation function as a function of $r$ from for three different temperatures and $N=1600$, plotted in the same way as Fig.~\ref{fig:dens}.}
  \label{fig:corr}
\end{figure*}

\subsection{Correlations and radial structure}

To get a more detailed statistical description of the configurations at different temperatures, we investigate the expectation values $\rho$ of the number of vertices at a distance $r$ from a vertex, in Fig.~\ref{fig:dens}. Panels (a) and (b) show the curves below, and at the maximum diameter.  For these curves, the radial density decreases exponentially, which, we believe, only a fragmentation of the networks can explain. In the $T\rightarrow\infty$ limit, the radial density curve looks peaked, as expected in random graphs~\cite{havlin:rad}, and similar to observations in large connected networks~\cite{our:rad}. 

In the \textit{XY} model on two-dimensional lattices, one of the central observations is that the spin correlations decay algebraically in the low temperature phase as opposed to an exponential decay for high temperatures. In Fig.~\ref{fig:corr}, we graph the correlation function
\begin{equation}
C(r) = \left|\left\langle e^{i((\theta_i-\theta_j)}\right\rangle_r \right|,
\end{equation}
where $\langle\;\cdot\;\rangle_r$ denote an average over vertex pairs separated by a graph distance $r$. If all vertex pairs at a certain distance have the same relative angular difference $C$ will be one. Comparing the four panels we note that (close to $r=0$) the correlations decay slowest for the lowest temperatures. This is the same behavior as in any classical spin model of statistical physics. But in panels (b) and (c) there is a second peak. We understand this by analogy to the Erd\H{o}s--R\'{e}nyi model --- if $2M>N$ most vertices of this model is connected into a ``giant component'', while the rest of the graph consists of small components. In our case, there are also disconnected subgraphs for larger temperatures (i.e., when the graph has a giant component). The non-giant components are sparser than the giant and should therefore be more volatile in structure. This leads to lower correlations at distances close to the diameter of the smaller components.

\section{Summary}

We have investigated the \textit{YX} model --- the \textit{XY} model where the connections, not the spins, are updated --- and found a complex pattern of size scaling. There are, for the sizes we study, three regimes: one regime of fragmented configurations dominating at low temperatures, an intermediate-temperature regime dominated by stretched-out networks, and a high temperature, disordered region. The different quantities we investigate, all capturing different aspects of the network structure, scale to zero with different exponents. Thus the borders between the regimes also scale to zero temperature. For example, we measure the diameter (quantifying how elongated the largest connected component is) and the size of the largest and second largest components. The peak in diameter scales to $T=0$ like $\sim N^{-1.52\pm 0.02}$ while the peak of the size of the second largest component scale to $T=0$ like $\sim N^{-1.44\pm 0.02}$. This multiscaling implies that the picture of three different regions will change as the sizes increase beyond the ones we sample --- e.g., the peak of the diameter will, for very large sizes, not coincide with the peak of the size of the second largest component.

We do not derive the \textit{YX} model as a model of some specific system in nature or society, but can we  \textit{a posteriori} identify such systems? Closest, we believe, is to interpret the \textit{YX} model as describing a social system where the spins represent opinions. As we have described the model, it does not describe the time evolution of the system. As most interesting questions about social systems regard their evolution, their models need to be dynamic. Dynamics can easily be added, most easily by taking the time evolution of e.g.\ Metropolis Monte Carlo sampling~\cite{mejn:statmech}. Whether or not this is reasonable depends on the system to be modeled. A condition on the subject system is that the social ties evolve faster than the opinions (as in models of social segregation~\cite{schelling:seg} or in some limits of coevolution models of networks and opinions~\cite{our:coevo,zanet:coevo}).

In the \textit{YX} model, the role of the $N\rightarrow\infty$ limit is, judging from our observations, that regime of fragmented states disappear. There are apparently no emergent singularities. On the other hand, the finite-size scaling shows a complex behavior with different scaling exponents. We cannot rule out a scenario our scaling parameters converge as $N\rightarrow\infty$ and there is a unique parameter $\chi$ such that all quantities signal the elongated-configuration regime at a critical $TN^\chi$. This, we believe, would be the most likely scenario of a phase transition. But nothing in our results suggests this would happen. On the other hand, the size scaling itself is highly complex. Viewed in this way, we have a statistical physics model where the finite sizes are more interesting than the thermodynamic limit. Moreover, since many real systems (especially in interdisciplinary physics) have restricted sizes, we believe focusing on size scaling rather than extrapolating to infinite sizes is a fruitful future direction for the analysis of statistical-mechanics models.

\acknowledgments{
The authors thank Beom Jun Kim for constructive comments.
PH acknowledges support from the Swedish Foundation for Strategic Research. ZXW and PM acknowledge support from the Swedish Research Council.
}

\end{document}